\documentclass[peerreview]{IEEEtran}
\usepackage{graphicx}
\usepackage{color}
\usepackage{amsmath}
\usepackage{relsize}

\begin{document}

\title{An Analytic Approach to the Modeling of Multi-Junction Solar Cells} 

\author{Rune~Strandberg        
\thanks{R. Strandberg is with the Department of Engineering Sciences, University of Agder, P.O. Box 509, NO-4898 Grimstad, Norway (e-mail: runes@uia.no).}}

\maketitle 

\begin{abstract}
Analytic expressions for the $JV$-characteristics of three types of multi-junction configurations are derived. From these, expressions for the short circuit current, open circuit voltage and voltage at the maximum power point are found for multi-terminal devices and for series-connected tandem stacks. For voltage-matched devices, expressions for the optimal ratio of the number of bottom cells to the number of top cells are established. Luminescent coupling is incorporated throughout the article. It should be highlighted that the maximum power point of a series connected tandem stack is described, with good accuracy for all interesting band gap combinations, by a single analytic expression. Although the modeling is carried out assuming devices that operate at the radiative limit, it is suggested that the results may form a framework that can be further developed and adapted to describe non-ideal devices as well.

\end{abstract}


\section{Introduction}
The role played by the ideal diode equation in our understanding of single junction solar cells can hardly be overrated. It reveals how a cell responds to illumination and bias and is the starting point for deriving expressions for central device characteristics such as power, temperature coefficients, open circuit voltage and short circuit current. Although advanced numerical device models exist for detailed simulation of single junction solar cells, the diode equation often captures the essence of the performance of such cells. So far, complete equivalents of the diode equation have not been published for multi-junction solar cells. Such devices are usually modeled by sets of nonlinear equations that are solved numerically to calculate various measures of device performance, as in Refs. \cite{Henry1980,DeVos1980, Marti1996, Brown2002, Bremner2008a, Geisz2015}. Exceptions include the work presented in Refs. \cite{Pusch2019, Friedman2013, Alam2016}. In the present work, equvivalents of the diode equation are derived for three types of multi-junction solar cells: multi-terminal devices, series-connected tandem stacks and voltage-matched tandem modules. These results are then used to establish expressions for device parameters like the short circuit current density, open circuit voltage and maximum power point for the first two device types whereas expressions for the optimal number of cells in the top and bottom layer are derived for the latter type. There are two important differences between Ref. \cite{Pusch2019} and the present work in the part on series-connected tandem cells: 1) In this work, the $JV$-characteristic of the entire device is derived, whereas Ref. \cite{Pusch2019} derives it for the top and bottom cells separately. 2) In Ref. \cite{Pusch2019}, different expressions must be used to calculate the maximum power point, depending on the degree of current-mismatch between the top and bottom cells. This work presents a single expression which can be applied to all useful combinations of band gaps. Another article that develops analytic $JV$-characteristics of the individual cells in a series-connected tandem stack is Ref. \cite{Friedman2013}. There the authors also developed expressions for some device parameters to allow better interpretation of experimental results. The present work represents a step forward in painting the full picture. In Ref. \cite{Alam2016} the authors developed an analytic model that approximates the efficiency limits of series-connected multi-junction devices. That model can be used to analyze stacks with several cells, whereas the present work only treats two series connected cells. The work in Ref. \cite{Alam2016} is based on approximations that makes it less accurate than the present model, however, and the former can only be applied to certain combinations of band gaps, i.e. band gaps which assures that the generation currents throughout the stack are equal. 

 The next few paragraphs give an introduction to the diode equation. Readers who are well acquainted with this may resume reading after Eq. (\ref{eq:SJPMPP}). The classic diode equation can be derived from device physics, thermodynamics or from a detailed balance perspective \cite{Shockley1949, Parrott1992, Marti1997}. In the latter approach, the recombination current density for a cell emitting to a hemisphere, which is given by \cite{DeVos1981}

\begin{equation}
\label{eq:rec}
J_\mathrm{R}=q\frac{2\pi}{h^3c^2}\int_{E_\mathrm{g}}^{\infty}\frac{E^2}{\mathrm{e}^{\frac{E-qV}{kT}}-1}\,\mathrm{d}E,
\end{equation}
is approximated as
\begin{equation}
\label{eq:recapprox}
J_\mathrm{R}\approx J_0\,\mathrm{e}^{qV/kT}
\end{equation}
with 
\begin{equation}
\label{eq:J0}
J_0=q\frac{2\pi}{h^3c^2}\int_{E_\mathrm{g}}^{\infty}E^2\,\mathrm{e}^{\frac{-E}{kT}}\,\mathrm{d}E.
\end{equation}
In the expressions above, $q$ is the elementary charge, $h$ is Planck's constant, $c$ is the speed of light in vacuum, $k$ is Boltzmann's constant, $T$ is the device temperature, $E$ is the energy of the photons emitted by the cell, $E_\mathrm{g}$ is the band gap of the cell material and $V$ is the cell voltage. The integrals are taken over the photon energies that the cell is allowed to emit to the surroundings. This gives the current density delivered by the cell as a function of its voltage as
\begin{equation}
\label{eq:SJIV}
J=J_\mathrm{G}-J_0\,\mathrm{e}^{\frac{qV}{kT}},
\end{equation}
where $J_\mathrm{G}$ is the generation current density experienced by the cell. In the radiative limit, $J_\mathrm{G}$ is calculated from the flux of incoming photons in the energy interval where the cell is absorbing. Whenever $J_0$ is much smaller than $J_\mathrm{G}$, a condition met by any useful solar cell, the short circuit current density $J_\mathrm{sc}$ is practically equal to $J_\mathrm{G}$. From (\ref{eq:SJIV}) one can derive the expression
\begin{equation}
\label{eq:Voc}
V_\mathrm{oc}=\frac{kT}{q}\ln \left(\frac{J_\mathrm{G}}{J_0}\right)
\end{equation}
for the open circuit voltage after equating the $JV$-characteristic to zero. While finding the open circuit voltage from (\ref{eq:SJIV}) is straightforward and has been described in literature for decades, the corresponding expression for the voltage giving the maximum power density is a rather recent finding as it was published by Sergeev and Sablon in 2018 \cite{sergeev2018}. The power density of the cell as a function of the voltage is found by multiplying (\ref{eq:SJIV}) by $V$. The maximum power point can then be found by determining the peak of the resulting $PV$-characteristic. Using the Lambert W function, the voltage at the maximum power point is given by
\begin{equation}
\label{eq:SJVMPP}
V_\mathrm{mpp}=	\frac{kT}{q}\left[\mathrm{W}\left(\frac{J_\mathrm{G}}{J_0}\mathrm{e}\right)-1 \right], 
\end{equation}
which gives the maximum power density
\begin{equation}
\label{eq:SJPMPP}
P_\mathrm{mpp}=\frac{kT}{q}J_\mathrm{G}\left[\mathrm{W}\left(\frac{J_\mathrm{G}}{J_0}\mathrm{e}\right) + \frac{1}{\mathrm{W}\left(\frac{J_\mathrm{G}}{J_0}\mathrm{e}\right)} -2 \right].
\end{equation}
The parameter $\mathrm{e}$ is the base of the natural logarithm, i.e. $\mathrm{exp}(1)$. Using (\ref{eq:SJPMPP}) reproduces the 40.8 \% efficiency limit for single junction cells illuminated by light from a 6000 K black body which follows from the work of Shockley and Queisser \cite{Shockley1961}. This is slightly larger than the $40.74\,\%$ limit found when using Eq. (\ref{eq:rec}) for the recombination current density \cite{Bremner2008a}. The error involved in using Eq. (\ref{eq:recapprox}) gets smaller, and eventually becomes insignificant, with decreasing light intensity.

In the first of the following sections, expressions equivalent to Eqs. (\ref{eq:SJIV}), (\ref{eq:Voc}), (\ref{eq:SJVMPP}) and (\ref{eq:SJPMPP}), are derived for multi-terminal multi-junction devices. Series-connected tandem stacks are treated in section \ref{sec:CM}. This is the most extensive, and to the author's opinion most important, part of the article. Finally, in section \ref{sec:VM}, the $JV$-characteristic of voltage-matched devices is investigated and the optimal ratio of the number of top cells to the number of bottom cells is determined. Luminescent coupling is incorporated in all three sections.

When modeling multi-junction devices, it is possible to incorporate non-radiative recombination, interface reflection or reduced emissivity/absorptivity into the model as in Ref. \cite{Pusch2019}. The increased complexity in doing so can easily reduce the clarity of the present work. Therefore, a constant refractive index is assumed for the entire devices throughout this article. It is also assumed that the recombination is purely radiative, that the devices have spectral absorptivities of either 0 or 1, and that there is no reflection in the systems other than unavoidable total internal reflection and full reflection from the backside/bottom of the stack. The models to be presented may be refined and adapted to include more details when applied to particular cases in future work.

\section{Multi-terminal multi-junction devices}
\label{sec:multi}
To differentiate the parameters of different cells in a multi-junction stack, indices are used, with the top cell having the index 1. The generation current of cell $i$ is then denoted $J_{\mathrm{G},i}$ and calculated from the photon flux in the energy interval from the band gap energy of cell $i$ to the energy gap of cell $i-1$, except for the top cell which receives all photons above the top cell band gap. The recombination parameter $J_{0,i}$ of cell $i$ is calculated from (\ref{eq:J0}) with the lower integration limit set to the band gap energy of cell $i$ and the upper limit set to infinity, if $i=1$, or to the band gap energy of cell $i-1$ otherwise. The maximum power of multi-junction devices with individually operated cells and no radiative coupling can now easily be calculated by applying Eq. (\ref{eq:SJPMPP}) to each cell and summing the contribution from each of them.  When radiative coupling is taking place, the cells exchange photons with each other and cannot be treated individually. Instead, a set of equations must be solved to find the maximum power point of each cell. The amount of radiation transferred from one cell to the neighboring cells depends on the design and material properties of the device. With a constant refractive index $n_\mathrm{r}$ throughout the stack, the contribution to the $JV$-characteristic of cells $i$ and $i+1$ associated with the luminescence radiated from cell $i$ to cell $i+1$ can be expressed as
\begin{equation}
\label{eq:LCterms} 
n_\mathrm{r}^2 \sum_{j=1}^i J_{0,j} \mathrm{e}^{\frac{qV_i}{kT}},
\end{equation}
where $V_i$ is the voltage of cell $i$. The luminescence emitted from cell $i+1$ to cell $i$ is given by the same expression, but with the voltage $V_{i+1}$ of cell $i+1$ replacing $V_i$. Due to the exponential function appearing in Eq. (\ref{eq:J0}), terms with $j\neq i$ are negligible if the difference between the band gaps is larger than a few times $kT$. For band gaps larger than 0.6 eV and a device temperature of 300 K, the error introduced by neglecting all terms with $j<i$ is smaller than $1\,\%$ if the difference between the band gaps is at least $5kT$. To achieve the highest efficiency, the difference between the band gaps of neighboring cells in stacks with up to 8 cells should be larger than this \cite{Bremner2008a}. Consequently, terms with $j\neq i$ in (\ref{eq:LCterms}) are neglected in the following.

\begin{figure}[htbp]
\centering
\includegraphics[width=9cm]{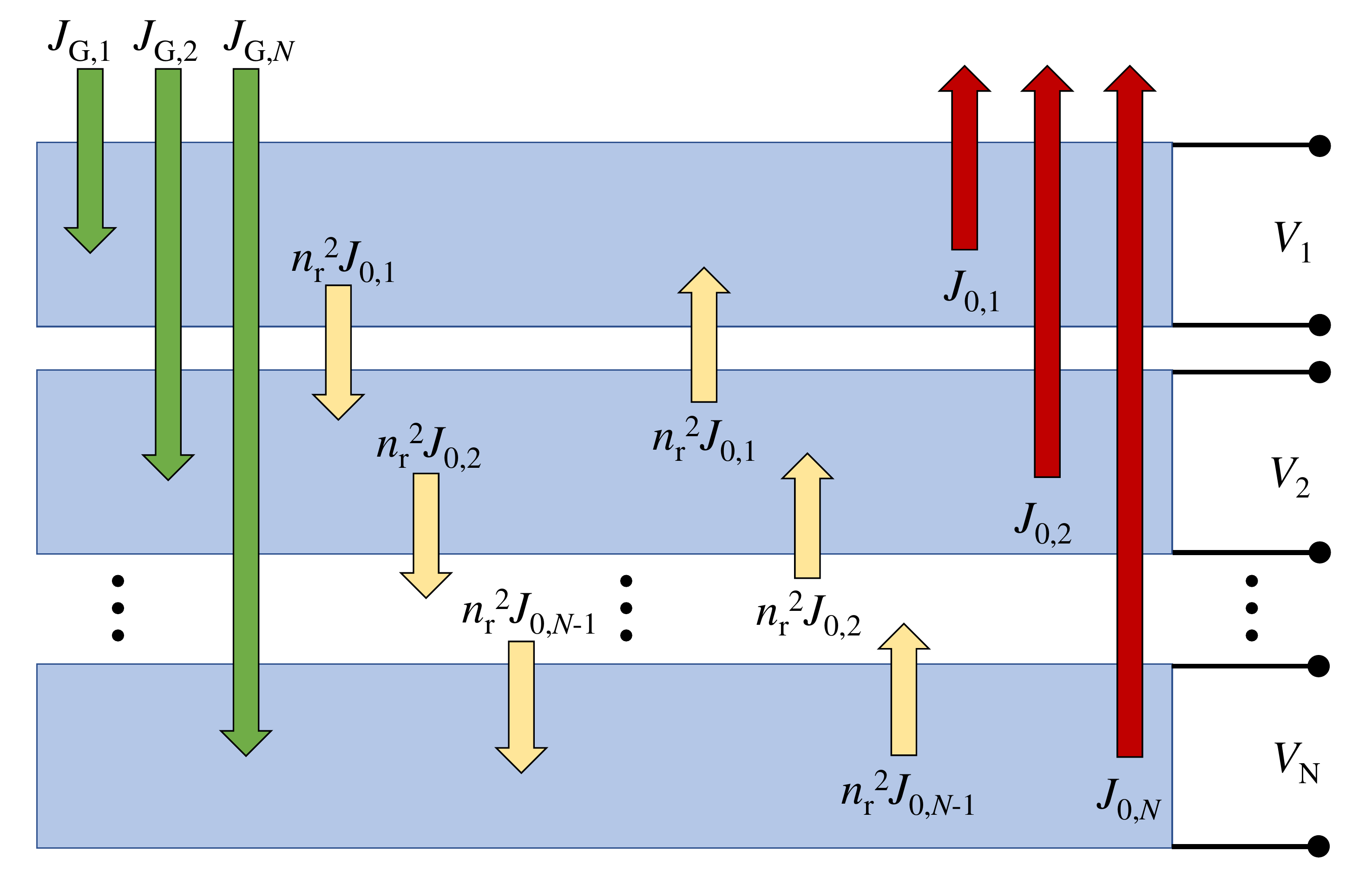}
\caption{Sketch illustrating the various routes of photon transport to, from and within a multi-terminal stack with $N$ cells. External illumination reaches each cell which is illustrated by the arrows marked with $J_{\mathrm{G},i}$. The remaining arrows illustrate photons emitted to the surroundings as well as luminescence transferred from one cell to another. These arrows are marked by the pre-exponential factors of the respective processes. \label{fig:multiterminal}}
\end{figure}

Figure \ref{fig:multiterminal} shows the different paths of photon transport in a multi-terminal stack. The arrows representing photons exchanged between cells or photons emitted to the surroundings are marked with the pre-exponential factors that is associated with the respective processes. For a cell which is not the top or bottom cell, the $JV$-characteristic becomes 
\begin{equation}
\label{eq:cellinstack}
J_i=J_{\mathrm{G},i}-\left[\left(1+n_\mathrm{r}^2\right)J_{0,i}+n_\mathrm{r}^2J_{0,i-1}\right]\mathrm{e}^{\frac{qV_i}{kT}}+n_\mathrm{r}^2J_{0,i-1}\mathrm{e}^{\frac{qV_{i-1}}{kT}}+n_\mathrm{r}^2J_{0,i}\mathrm{e}^{\frac{qV_{i+1}}{kT}}.
\end{equation}
This expression for the current density consists of four parts. The first part is simply the generation current associated with the absorption of external photons. The second part includes luminescence emitted by cell $i$ to the surroundings as well as to its two neighboring cells. The third term takes into account luminescence emitted to cell $i$ from the cell above, and the last term accounts for luminescence from the cell below. The top and bottom cells will have $JV$-characteristics similar to (\ref{eq:cellinstack}), but terms associated with transfer of photons to cells above the top cell or below the bottom cell will be zero.

The purpose of a multi-junction stack is to extract electrons excited by high-energy photons at a higher voltage than electrons excited by photons with lower energy. For stacks with a limited number of cells, the difference in voltage between neighboring cells should be at least a few times $kT/q$, reflecting the difference in band gaps. When this is the case, the number of photons emitted from cell $i+1$ to cell $i$ will be negligible compared to the number of photons reaching cell $i$ from other sources. The current density of a cell which is not the top cell, now becomes 
\begin{equation}
\label{eq:cellinstack2}
J_i=J_{\mathrm{G},i}-\left(1+n_\mathrm{r}^2\right)J_{0,i}\mathrm{e}^{\frac{qV_i}{kT}}+n_\mathrm{r}^2J_{0,i-1}\mathrm{e}^{\frac{qV_{i-1}}{kT}},
\end{equation}
where the second part has also been simplified because $J_{0,i}$ is typically much larger than $J_{0,i-1}$ as argued above. Since the current density is now independent of the voltage of the cell below, the performance of the stack can be calculated cell by cell, starting from the top of the stack. 

Using the result of Sergeev and Sablon \cite{sergeev2018}, the optimal voltage for the top cell then becomes 
\begin{equation}
\label{eq:Vmpp_top}
V_\mathrm{mpp,1} = \frac{kT}{q}\left[\mathrm{W}\left(Z_1\right)-1 \right],
\end{equation}
where
\begin{equation}
\label{eq:Zmpp_top}
Z_1 = \frac{J_\mathrm{G,1}}{(1+n_\mathrm{r}^2)J_{0,1}}\mathrm{e}.
\end{equation}
This gives the maximum power density 
\begin{equation}
\label{eq:Pmpp_top}
P_1=\frac{kT}{q}J_\mathrm{G,1}\left[ \mathrm{W}\left( Z_1 \right) + \frac{1}{\mathrm{W}\left( Z_1 \right)} -2 \right],
\end{equation}
For the second cell in the stack, assuming this is not the bottom cell, following the same procedure and taking into account luminescence from the top cell gives
\begin{equation}
\label{eq:Pmpp_2}
P_2=\frac{kT}{q}\left(J_\mathrm{G,2}+n_\mathrm{r}^2 J_{0,1}\mathrm{e}^{\frac{qV_\mathrm{1}}{kT}}\right)\left[ \mathrm{W}\left( Z_2 \right) + \frac{1}{\mathrm{W}\left( Z_2 \right)} -2 \right],
\end{equation}
where
\begin{equation}
\label{eq:Zmpp_2}
Z_2 = \frac{J_\mathrm{G,2}+n_\mathrm{r}^2J_{0,1}\mathrm{e}^{\frac{qV_\mathrm{1}}{kT}}}{(1+n_\mathrm{r}^2)J_{0,2}}\mathrm{e}.
\end{equation}
More general, in a stack with $N$ cells the optimal voltage for cell $i$ is given by
\begin{equation}
\label{eq:Vmpp_i}
V_\mathrm{mpp,i} = \frac{kT}{q}\left[\mathrm{W}\left(Z_i\right)-1 \right].
\end{equation}
For $1<i<N$
\begin{equation}
\label{eq:Zmpp_i}
Z_i = \frac{J_\mathrm{G,i}+n_\mathrm{r}^2J_{0,i-1}\mathrm{e}^{\frac{qV_\mathrm{i-1}}{kT}}}{(1+n_\mathrm{r}^2)J_{0,i}}\mathrm{e},
\end{equation}
and for the bottom cell
\begin{equation}
\label{eq:Zmpp_N}
Z_N = \frac{J_\mathrm{G,N}+n_\mathrm{r}^2J_{0,N-1}\mathrm{e}^{\frac{qV_\mathrm{N-1}}{kT}}}{J_{0,N}}\mathrm{e}.
\end{equation}
The maximum power density for cell $i$ where $i>1$ is now given by
\begin{equation}
\label{eq:Pmpp_i}
P_i=\frac{kT}{q}\left(J_\mathrm{G,i}+n_\mathrm{r}^2 J_{0,i-1}\mathrm{e}^{\frac{qV_\mathrm{i-1}}{kT}}\right)\left[ \mathrm{W}\left( Z_i \right) + \frac{1}{\mathrm{W}\left( Z_i \right)} -2 \right].
\end{equation}
The total power density of the entire stack is finally found by summing the contribution from all cells.

If all the cells in the stack are at their respective maximum power points, we have
\begin{equation}
n_\mathrm{r}^2J_{0,i-1}\mathrm{e}^{\frac{qV_\mathrm{i-1}}{kT}} = \mathlarger{\sum}_{j=1}^{i-1} \frac{\left(\frac{n_\mathrm{r}^2}{1+n_\mathrm{r}^2}\right)^{i-j}}{\prod_{k=j}^{i-j}\mathrm{W}(Z_k)}J_{\mathrm{G},j},
\end{equation}
which can be verified by substituting for $V_{i-1}$ in Eq. (\ref{eq:Zmpp_i}) using Eqs. (\ref{eq:Zmpp_top}) and (\ref{eq:Vmpp_i}) as well as (\ref{eq:Zmpp_i}) itself.

Starting with the current density (\ref{eq:cellinstack}), it is also possible to establish expressions for the open circuit voltage and short circuit current density. Realizing that $J_\mathrm{G,i}$ is much larger than any of the parameters $J_{0,j}$ for any meaningful combination of band gaps, the short circuit current density of cell $i$ is simply 
\begin{equation}
\label{eq:SC4T}
J_{i,\mathrm{sc}}=J_{\mathrm{G},i}+n_\mathrm{r}^2J_{i-1,i}\mathrm{e}^{\frac{qV_{i-1}}{kT}}+n_\mathrm{r}^2J_{i+1,i}\mathrm{e}^{\frac{qV_{i+1}}{kT}}
\end{equation}
where any term with $i-1=0$ or $i+1=N+1$ are zero. If both neighboring cells are also operating at short circuit the two last terms disappear and the short circuit current is simply equal to the generation current of the cell in question. 

The open circuit voltage of cell $i$ is found by setting $J_i=0$ and solving for $V_i$. By neglecting luminescence from the cell below it becomes
\begin{equation}
\label{eq:VOC4T1}
V_{1,\mathrm{oc}}=\frac{kT}{q}\mathrm{ln}\left(\frac{J_{\mathrm{G},1}}{\left(1+n_\mathrm{r}^2\right)J_{0,1}}\right)
\end{equation} 
for the top cell,
\begin{equation}
\label{eq:VOC4Ti}
V_{i,\mathrm{oc}}=\frac{kT}{q}\mathrm{ln}\left(\frac{J_{\mathrm{G},i}+n_\mathrm{r}^2J_{0,i-1}\mathrm{e}^{\frac{qV_{i-1,}}{kT}}}{\left(1+n_\mathrm{r}^2\right)J_{0,i}}\right)
\end{equation}
for any cell not at the very top or bottom of the stack, and 
\begin{equation}
\label{eq:VOC4TN}
V_{N,\mathrm{oc}}=\frac{kT}{q}\mathrm{ln}\left(\frac{J_{\mathrm{G},N}+n_\mathrm{r}^2J_{0,N-1}\mathrm{e}^{\frac{qV_{N-1}}{kT}}}{J_{0,N}}\right).
\end{equation}
for the bottom cell. If all cells in the stack are at open circuit, Eq. (\ref{eq:VOC4Ti}) can be rewritten as
\begin{equation}
\label{eq:VOC4Tigen}
V_{i,\mathrm{oc}}=\frac{kT}{q}\mathrm{ln}\left(\frac{\sum_{j=1}^{i}\left(\frac{n_\mathrm{r}^2}{1+n_\mathrm{r}^2} \right)^{i-j}J_{\mathrm{G},j}}{\left(1+n_\mathrm{r}^2\right)J_{0,i}}\right)
\end{equation}
which is also valid for the top cell. The expression for the bottom cell would be similar, but the refractive index does not appear in the denominator of the argument of the logarithm.

\section{Series-connected tandem stacks}
\label{sec:CM}
\begin{figure}[htbp]
\centering
\includegraphics[width=9cm]{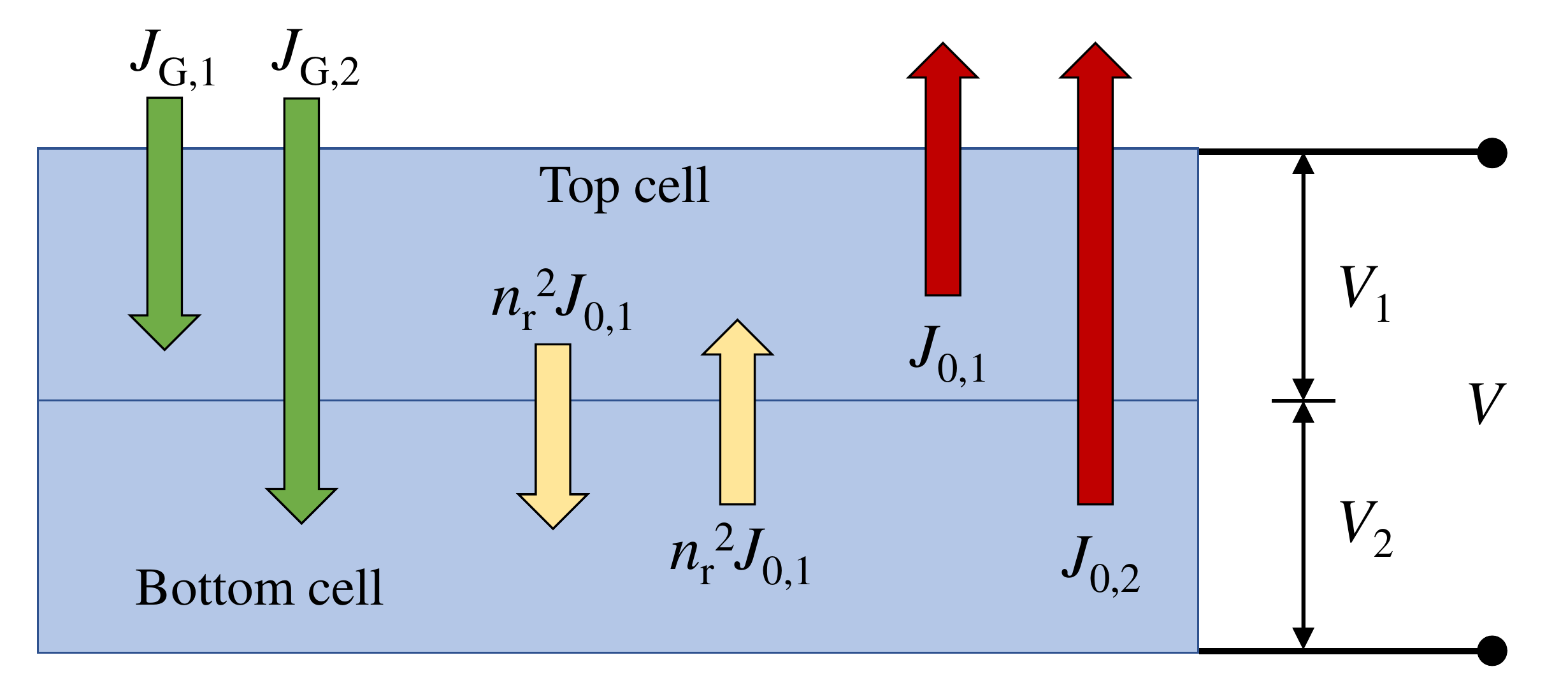}
\caption{Sketch illustrating the various routes of photon transport within a series-connected tandem stack. External illumination reaches each cell which is illustrated by the arrows marked with $J_{\mathrm{G},1}$ and $J_{\mathrm{G},1}$. The remaining arrows illustrate photons emitted to the surroundings as well as luminescence transferred between the cells. These arrows are marked by the pre-exponential factors of the respective processes.  \label{fig:seriesconnected}}
\end{figure}

In a series-connected tandem stack, two cells are placed on top of each other and the current is extracted from two terminals, one connected to each cell, as sketched in Fig. \ref{fig:seriesconnected}. The series-connection subjects the device to the constraint $J_1=J_2 \equiv J$. The voltages of the cells cannot be selected independently, but will take values that assures the fulfillment of this constraint. The sum of the voltages set up by the two cells equals the voltage $V$ between the two terminals, i.e. $V_1+V_2=V$. The relation between the current density and cell voltage for the individual cells in the stack bear resemblance to those of the cells in a multi-terminal stack. Analogous to Eq. (\ref{eq:cellinstack}), the current densities 
\begin{equation}
\label{eq:topcellCM}
J_1=J_{\mathrm{G},1}-\left(1+n_\mathrm{r}^2\right)J_{0,1}\mathrm{e}^{\frac{qV_1}{kT}}+n_\mathrm{r}^2 J_{0,1}\mathrm{e}^{\frac{qV_2}{kT}},
\end{equation}
for the top cell, and  
\begin{equation}
\label{eq:bottomcellCM}
J_2=J_{\mathrm{G},2}-\left(J_{0,2}+n_\mathrm{r}^2J_{0,1}\right)\mathrm{e}^{\frac{qV_2}{kT}}+n_\mathrm{r}^2 J_{0,1}\mathrm{e}^{\frac{qV_1}{kT}},
\end{equation}
for the bottom cell, can be established. Note that the validity of these expressions do not depend on any assumptions about the difference between the band gaps of the two cells.
To find the $JV$-characteristic for the tandem stack the current density has to be expressed as a function of $V$ rather than $V_1$ and $V_2$. Before proceeding with that derivation, it is useful to define a few new parameters. As will become clear later, the parameters
\begin{equation}
\label{eq:transfer1}
T_{1\rightarrow 2}=\frac{n_\mathrm{r}^2}{1+2n_\mathrm{r}^2}
\end{equation}
and
\begin{equation}
\label{eq:transfer2}
T_{2\rightarrow 1}=\frac{n_\mathrm{r}^2 J_{0,1}}{J_{0,2}+2n_\mathrm{r}^2 J_{0,1}}
\end{equation}
are transfer coefficients which describes the degree of radiative coupling between the two cells. To shorten expressions that appear in the following, the two linear combinations

\begin{equation}
\label{eq:Tplus}
T^+=T_{1\rightarrow 2}+T_{2\rightarrow 1}
\end{equation}
and
\begin{equation}
\label{eq:Tminus}
T^-=T_{1\rightarrow 2}-T_{2\rightarrow 1}
\end{equation}
of the transfer coefficients come in handy. Finally, defining the parameter
\begin{equation}
\label{eq:J0CM}
\tilde{J}_0^2=\left(1+2n_\mathrm{r}^2\right)\left(J_{0,2}+2n_\mathrm{r}^2 J_{0,1}\right) J_{0,1}
\end{equation}
is also convenient.

Defining 
\begin{equation*}
\Delta J_\mathrm{G} = J_{\mathrm{G},1} - J_{\mathrm{G},2}
\end{equation*}
and applying $J_1=J_2$, gives
\begin{equation}
\label{eq:CM}
\Delta J_\mathrm{G} = \left(1+2n_\mathrm{r}^2\right)J_{0,1}\mathrm{e}^{\frac{qV_1}{kT}}-\left(J_{0,2}+2n_\mathrm{r}^2J_{0,1} \right)\mathrm{e}^{\frac{qV_2}{kT}}.
\end{equation}
Using this result in combination with Eqs. (\ref{eq:topcellCM}) and (\ref{eq:bottomcellCM}), gives the $JV$-characteristics of the top and bottom cells, as a function of the respective cell voltages only, as
\begin{equation}
\label{eq:JVtop}
J_1=J_{\mathrm{G},1}-T_{1\rightarrow 2}\Delta J - \left(1-T^+\right)\left(1+2n_\mathrm{r}^2\right)J_{0,1}\mathrm{e}^\frac{qV_1}{kT}
\end{equation}
and
\begin{equation}
\label{eq:JVbot}
J_2=J_{\mathrm{G},2}+T_{2\rightarrow 1}\Delta J - \left(1-T^+\right)\left(J_{0,2}+2n_\mathrm{r}^2J_{0,1}\right)\mathrm{e}^\frac{qV_2}{kT}.
\end{equation}

To proceed towards the $JV$-characteristic of the tandem stack, either $V_1$ or $V_2$ can be eliminated from Eq. (\ref{eq:CM}) by multiplying that equation by $\mathrm{exp}\left(qV_2/(kT) \right)$ or $\mathrm{exp}\left(qV_1/(kT) \right)$, respectively, before using $V=V_1+V_2$. The resulting expressions are second order equations of either $\mathrm{exp}\left(qV_2/(kT) \right)$ or $\mathrm{exp}\left(qV_1/(kT) \right)$, with the solutions
\begin{equation}
\label{eq:V1}
\mathrm{e}^{\frac{qV_1}{kT}}=\frac{\frac{1}{2}\Delta J_\mathrm{G}+\sqrt{\tfrac{1}{4}\Delta J_\mathrm{G}^2 + \tilde{J}_0^2\mathrm{e}^{\frac{qV}{kT}}}}{\left(1+2n_\mathrm{r}^2\right)J_{0,1}}
\end{equation}
and
\begin{equation}
\label{eq:V2}
\mathrm{e}^{\frac{qV_2}{kT}}=\frac{-\frac{1}{2}\Delta J_\mathrm{G}+\sqrt{\tfrac{1}{4}\Delta J_\mathrm{G}^2 + \tilde{J}_0^2\mathrm{e}^{\frac{qV}{kT}}}}{\left(J_{0,2}+2n_\mathrm{r}^2 J_{0,1}\right)}.
\end{equation}
Inserting (\ref{eq:V1}) and (\ref{eq:V2}) into either (\ref{eq:topcellCM}) or (\ref{eq:bottomcellCM}) now gives the $JV$-characteristic of the stack as

\begin{equation}
\label{eq:IVCMRC}
J=\tfrac{1}{2}\left(J_\mathrm{G,1}+J_\mathrm{G,2}\right)+\tfrac{1}{2}T^-\Delta J_\mathrm{G}-\left(1-T^+\right)\sqrt{\tfrac{1}{4}\Delta J_\mathrm{G}^2+\tilde{J}_0^2\mathrm{e}^{\frac{qV}{kT}}}.
\end{equation}
If there is no radiative coupling between the two cells, the expression for the current is given by the simpler
\begin{equation}
\label{eq:IVCM}
J=\tfrac{1}{2}\left(J_\mathrm{G,1}+J_\mathrm{G,2}\right)-\sqrt{\tfrac{1}{4}\Delta J_\mathrm{G}^2+J_{0,1}J_{0,2}\mathrm{e}^{\frac{qV}{kT}}}.
\end{equation}
The first part of this simpler $JV$-characteristic is a pure generation term, whereas the square root envelops a recombination term, containing the voltage, and a penalty term in case of a mismatch between the generation current densities of the two cells. Readers familiar with the $JV$-characteristic of intermediate band solar cells (IBSCs) \cite{Strandberg2017b} will recognize (\ref{eq:IVCM}) as a simpler form of the JV-characteristic found for said concept. In Ref. \cite{Wilkins2020}, Wilkins et al. found that such a simple analytic $JV$-characteristic captured the essence of the behavior of IBSCs predicted by a far more comprehensive numerical device model. This gives reason to believe that modified versions of Eqs. (\ref{eq:IVCMRC}) and (\ref{eq:IVCM}) may capture the essence of the behavior of non-ideal series-connected tandem cells if appropriate values are assigned to its parameters. It also worth pointing out that the radiative coupling in a tandem stack has much of the same impact on the form of the $JV$-characteristic as overlapping absorption coefficients have on the $JV$-characteristic of intermediate band solar cells \cite{Strandberg2017a}.

\subsection{Short circuit current density}
\label{sec:Jsc}
From Eq. (\ref{eq:IVCMRC}) it is straightforward to derive expressions for the short circuit current density. For any useful configurations of band gaps, $\tilde{J}_0$ will be orders of magnitude smaller than $J_\mathrm{G,1}$ and $J_\mathrm{G,2}$. The term containing $\tilde{J}_0$ can therefore be neglected when the voltage is set to zero. The remaining terms give

\begin{equation}
\label{eq:Jsc}
J_\mathrm{sc}=\tfrac{1}{2}\left[ \left(1+T^-\right)J_\mathrm{G,1}+\left(1-T^-\right)J_\mathrm{G,2}-\left(1-T^+\right)\vert \Delta J_\mathrm{G}\vert \right].
\end{equation}
If $J_{G1}>J_{G2}$, $J_\mathrm{sc}$ may be written

\begin{equation}
\label{eq:Jscpos}
J_\mathrm{sc}=J_\mathrm{G,2}+T_{1\rightarrow 2}\Delta J_\mathrm{G},
\end{equation}
whereas if $J_{G1}<J_{G2}$, it is given by

\begin{equation}
\label{eq:Jscneg}
J_\mathrm{sc}=J_\mathrm{G,1}-T_{2\rightarrow 1}\Delta J_\mathrm{G}.
\end{equation}
From these last two expressions for $J_\mathrm{sc}$, the interpretation of the transfer coefficients becomes clear. They quantify the fraction of the current density-mismatch $\Delta J$ which is transferred to the cell that acts as a bottleneck at short circuit. $T_{1\rightarrow 2}$ takes the value 1/3 if the refractive index equals 1. In this case one third of the current-mismatch is transferred from the top cell to the bottom cell. Another third is emitted from the top cell to the surroundings while the last third is kept in the top cell to maintain the current-matching. Since the radiative coupling increases with the refractive index, so does the value of $T_{1\rightarrow 2}$ up to a hypothetical maximum of 1/2 for an infinitely large refractive index. Normally, $J_{0,1}$ is orders of magnitude smaller than $J_{0,2}$. The value of $T_{2\rightarrow 1}$ will thus be negligible in most cases. For realistic values of the refractive index, exceptions to this will only be found in cases where the difference between the two band gaps is small. As the two band gaps approach each other, $T_{2\rightarrow 1}$ will also approach a maximum value of 1/2 since $J_{0,2}$ will then approach zero.

\subsection{Open circuit voltage}
\label{sec:Voc}
An expression for the open circuit voltage of the stack is most easily found by summing the open circuit voltages of the two cells found from Eqs. (\ref{eq:JVtop}) and (\ref{eq:JVbot}). This gives
\begin{equation}
\label{eq:Voc2}
V_\mathrm{oc}=\frac{kT}{q}\mathrm{ln}\left[ \frac{ \left(J_\mathrm{G1}-T_{2\rightarrow 1} \Delta J_\mathrm{G}\right)\left( J_\mathrm{G2}+T_{1\rightarrow 2} \Delta J_\mathrm{G} \right)}{\left(1-T^+ \right)^2\tilde{J}_0^2} \right].
\end{equation}
It is worth pointing out that the open circuit voltage does not depend on $\Delta J$ when there is no radiative coupling. Since no current is flowing through the stack, the cells can set up their individual open circuit voltages regardless of the current-matching. When radiative coupling is taking place it is satisfying that the numerator in Eq. (\ref{eq:Voc2}) suggests that the generation current in the bottom cell is enhanced by some photons transferred from the top cell when $\Delta J_\mathrm{G}$ is positive. It may appear peculiar, however, that the bottom cell seems to lose a considerable fraction of its generation current when $\Delta J_\mathrm{G}$ is negative. This is seemingly inconsistent with the fact that transfer of photons from the bottom cell to the top cell should be small in most cases. The explanation for this apparent contradiction is found in the factor $1-T^+$ in the denominator, which can be seen after a couple of further approximations. When $J_\mathrm{0,1}<<J_\mathrm{0,2}$, $1-T^+$ can be approximated by $(1+n_\mathrm{r}^2)/(1+2n_\mathrm{r}^2)$. For the same conditions, we also have

\begin{equation}
\label{eq:Tapprox}
\left(1-T^+\right)^2\tilde{J}_0^2 \approx \frac{\left(1+n_\mathrm{r}^2 \right)^2}{1+2n_\mathrm{r}^2}J_\mathrm{0,1}J_\mathrm{0,2}.
\end{equation}
Inserting this approximation into Eq. (\ref{eq:Voc2}) while approximating $T_{2\rightarrow 1}$ to zero allows us to write the open circuit voltage as
\begin{equation}
\label{eq:Voc3}
V_\mathrm{oc}\approx\frac{kT}{q}\mathrm{ln}\left[ \frac{ J_\mathrm{G1}\left( J_\mathrm{G2}+\frac{n_\mathrm{r}^2}{1+n_\mathrm{r}^2}J_\mathrm{G1} \right)}{\left(1+n_\mathrm{r}^2\right)J_{0,1}J_{0,2}} \right].
\end{equation}
This is a satisfying result because it is now clear that the generation current of the bottom cell is enhanced by photons transferred from the top cell regardless of the polarity of $\Delta J_\mathrm{G}$. In the denominator, the enhanced recombination in the top cell, due to the radiative coupling, is taken into account by the factor $1+n_\mathrm{r}^2$.

Using the same approximations also allows the open circuit voltages of the individual cells to be simplified. We get

\begin{equation}
\label{eq:Voct}
V_\mathrm{oc,1}=\frac{kT}{q}\mathrm{ln}\left[ \frac{J_\mathrm{G1}-T_{2\rightarrow 1} \Delta J_\mathrm{G}}{\left(1-T^+\right)\left(1+2n_\mathrm{r}^2 \right)J_{0,1}} \right]\approx\frac{kT}{q}\mathrm{ln}\left[ \frac{ J_\mathrm{G1}}{\left(1+n_\mathrm{r}^2\right)J_{0,1}} \right],
\end{equation}
for the top cell, and 

\begin{equation}
\label{eq:Vocb}
V_\mathrm{oc,2}=\frac{kT}{q}\mathrm{ln}\left[ \frac{J_\mathrm{G2}+T_{1\rightarrow 2} \Delta J_\mathrm{G}}{\left(1-T^+\right)\left(J_{0,2}+2n_\mathrm{r}^2 J_{0,1}\right)} \right]\approx\frac{kT}{q}\mathrm{ln}\left[ \frac{\left( J_\mathrm{G2}+\frac{n_\mathrm{r}^2}{1+n_\mathrm{r}^2}J_\mathrm{G1} \right)}{J_{0,2}} \right],
\end{equation}
for the bottom cell. The latter approximation is equivalent to Eq. (12a) in Ref. \cite{Friedman2013}. It is claimed in Ref. \cite{Friedman2013} that the open circuit voltage of the top cell is not affected by the luminescent coupling. This is not strictly correct, since the value of the denominator in Eq. \ref{eq:Voct} depends on the presence or absence of such coupling. Comparing (\ref{eq:Voct}) and (\ref{eq:Vocb}) to (\ref{eq:VOC4T1}) and (\ref{eq:VOC4TN}) shows that the open circuit voltages of the two individual cells in a series-connected tandem stack is equal to those of the corresponding cells in a four-terminal stack of two independently operated cells. Since no current is flowing through the cell as open circuit, the open circuit voltage does not depend on the wiring of the cells.

\subsection{Maximum power point}
\label{sec:mpp}
Following the procedure of Ref. \cite{sergeev2018} allows an exact expression for the maximum power point to be derived if the two cells in the stack are current-matched. In this case, the power density delivered by the tandem stack is given by

\begin{equation}
\label{eq:PCM}
P=\tfrac{1}{2}V\left(J_\mathrm{G,1}+J_\mathrm{G,2} \right)-\left(1-T^+\right)\tilde{J}_0 V\mathrm{e}^{\frac{qV}{2kT}}.
\end{equation}
Differentiating (\ref{eq:PCM}) with respect to $V$, equaling the result to zero and reorganizing yields

\begin{equation}
\label{eq:SolveCM}
\left(1+\frac{qV_\mathrm{mpp}}{2kT} \right)\mathrm{e}^{\frac{qV_\mathrm{mpp}}{2kT}+1}=\mathrm{e}\sqrt{\frac{J_\mathrm{G1}J_\mathrm{G2}}{\left(1-T^+ \right)^2\tilde{J}_0^2}},
\end{equation}
which has the solution

\begin{equation}
\label{eq:Vmppcm}
V_\mathrm{mpp}=\frac{2kT}{q}\left[\mathrm{W}\left(\mathrm{e}\sqrt{\frac{J_\mathrm{G1}J_\mathrm{G2}}{\left(1-T^+ \right)^2\tilde{J}_0^2}} \right)-1 \right].
\end{equation}
It has been used that $\tfrac{1}{2}\left( J_\mathrm{G1}+J_\mathrm{G2} \right)=\sqrt{J_\mathrm{G1}J_\mathrm{G2}}$ when $\Delta J_\mathrm{G}=0$. Multiplying Eq. (\ref{eq:IVCMRC}) by the voltage of the stack and inserting (\ref{eq:Vmppcm}) gives the power density
\begin{equation}
\label{eq:PmppCM}
P_\mathrm{mpp}=\frac{kT}{q}\left(J_\mathrm{G1}+J_\mathrm{G2}\right)\left[ \mathrm{W}\left(\sqrt{\tilde{Z}_1\tilde{Z}_2}\right)+\mathrm{W}\left(\sqrt{\tilde{Z}_1\tilde{Z}_2}\right)^{-1}-2\right]
\end{equation}
for a current-matched stack. $\tilde{Z}_1$ and $\tilde{Z}_2$ are defined as 
\begin{equation}
\tilde{Z}_1=\mathrm{e}\frac{J_\mathrm{G,1}-T_{2\rightarrow 1}\Delta J_\mathrm{G}}{(1-T^+)\left(1+2n_\mathrm{r}^2 \right)J_{0,1}}
\end{equation}
and
\begin{equation}
\tilde{Z}_2=\mathrm{e}\frac{J_\mathrm{G,2}+T_{1\rightarrow 2}\Delta J_\mathrm{G}}{(1-T^+)\left(J_{0,2}+2n_\mathrm{r}^2 J_{0,1}\right)}.
\end{equation}
It should be noted that $\sqrt{\tilde{Z}_1\tilde{Z}_2}=\exp\left(qV_\mathrm{oc}/2kT+1\right)$, which relates the maximum power point to the open circuit voltage .

\begin{figure}[htbp]
\centering
\includegraphics[width=9cm]{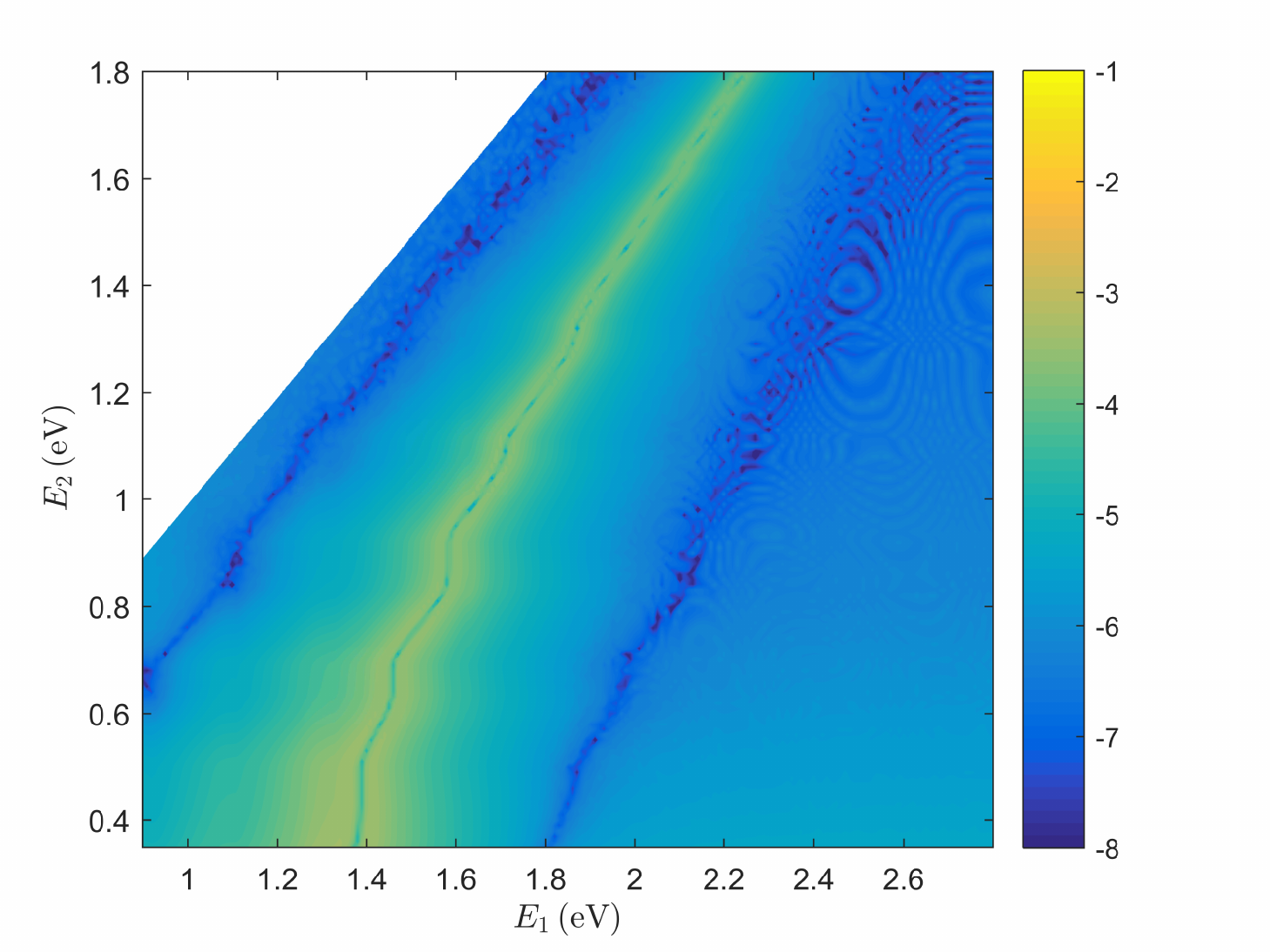}
\caption{The logarithm of the relative error of the maximum power density for stacks without luminescent coupling calculated using Eq. (\ref{eq:VmppCMTS}.) The AM1.5G spectrum has been used. The largest value of -3.39 is found for a device with band gaps of 0.35 and 1.34 eV. \label{fig:error}}
\end{figure}

Eq. (\ref{eq:Vmppcm}) complements the work of Pusch et al. \cite{Pusch2019} where expressions for the maximum power point was found for current-mismatched stacks. By combining the results for the maximum power point of current-matched and current-mismatched stacks, it is possible to construct a single approximate expression for the $V_\mathrm{mpp}$ which gives the maximum power density with good accuracy for a wide range of band gap combinations, including all combinations that may yield high efficiency under the AM1.5G spectrum. As a measure of the error in the calculation of the maximum power $P$, the logarithm of the relative error, $\mathrm{log}_{10}(\vert P_\mathrm{DB}-P\vert/P_\mathrm{DB})$, where $P_\mathrm{DB}$ is the power density calculated by a numerical detailed balance model, will be used. Calculating the maximum power density using the approximation
\begin{equation}
\label{eq:VmppCMTS}
\frac{qV_\mathrm{mpp}}{kT}+1\approx\mathrm{W}\left[\frac{\mathrm{e}J_{\mathrm{sc}}^2}{(1-T^+)^2\tilde{J}_0^2}\left((1-T^+)\frac{\vert\Delta J_\mathrm{G}\vert}{J_\mathrm{sc}}+\frac{2}{\mathrm{W}\left(\mathrm{e}^{\frac{qV_\mathrm{oc}}{2kT}+1}\right)}\right)\right]
\end{equation}
leads to the error shown in Fig. \ref{fig:error} when $n_\mathrm{r}$ is set to zero, that is, when radiative coupling is not present. The largest logarithm of the relative error shown in the figure equals -3.39, which corresponds to a deviation of $0.04\,\%$ from the detailed balance power density.  An explanation of how to arrive at this approximation for $V_\mathrm{mpp}$, as well as some more material on the accuracy of this and other expressions for the maximum power point, is included as an appendix.

\section{Voltage-matched devices}
\label{sec:VM}

\begin{figure}[htbp]
\centering
\includegraphics[width=9cm]{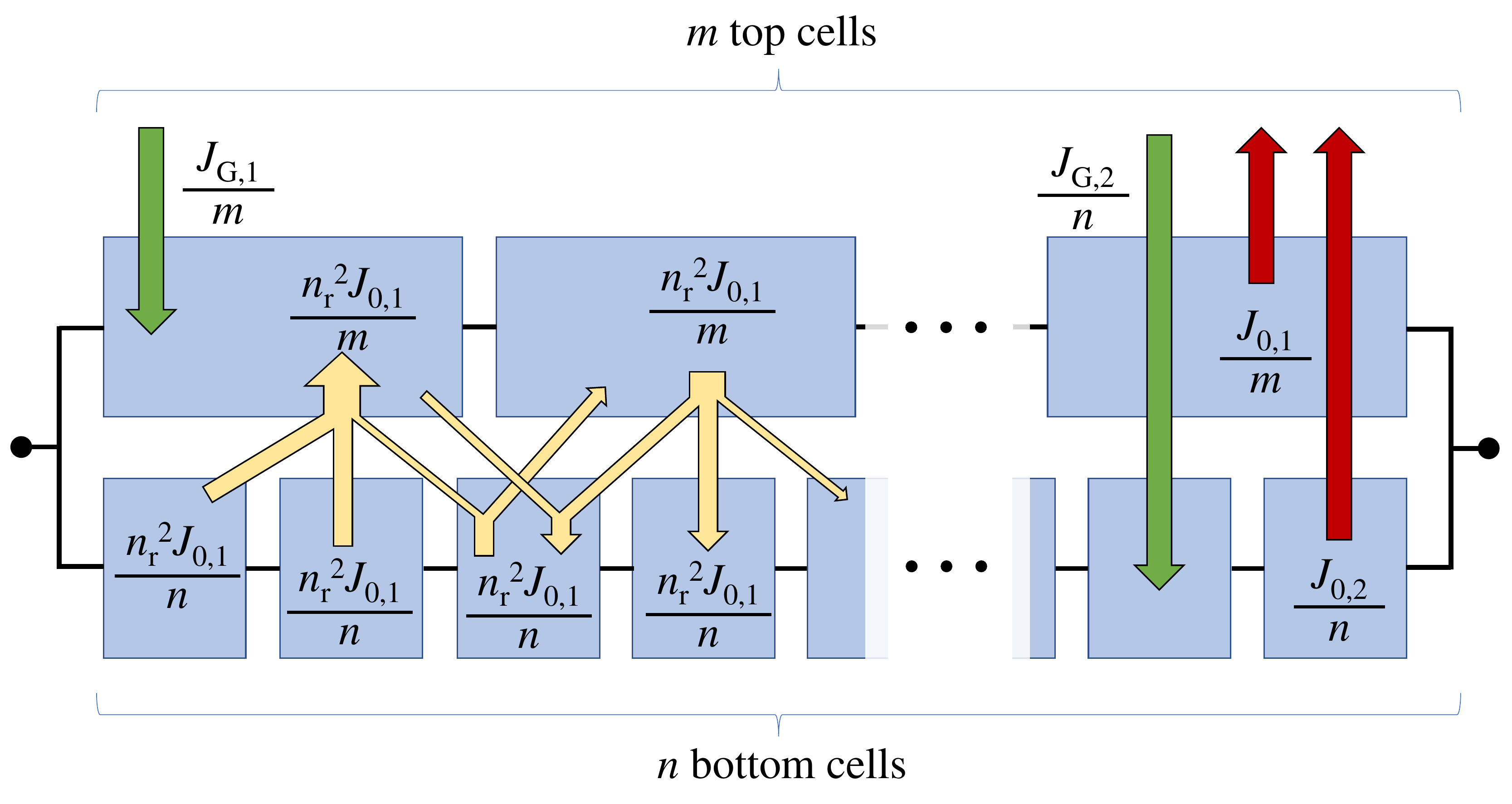}
\caption{A sketch of the routes of photon transport to, from and within an area de-coupled tandem device. Since the cells in each layer are equal and operate at the same voltage any cell will receive a flux of luminescent photons which is equal to what they would receive if the neighboring layer consisted of a single cell. The arrows indicating fluxes of incoming photons are marked by their contribution to the $JV$-characteristic of the device. The remaining arrows are marked by the pre-exponential factors associated with the respective processes in this $JV$-characteristic.  \label{fig:areadecoupled}}
\end{figure}

Layers of top and bottom cells can be voltage matched by using different string lengths in the top and bottom layers \cite{Lentine2014,Strandberg2015}. In the case of area de-coupling, a layer of $m$ series-connected top cells is connected in parallel with a layer of $n$ series-connected bottom cells. One cell in the top layer has an area $A_1$ while one cell in the bottom layer has an area $A_2$. The two layers cover the same total area $A$ and the spacing between the cells within each layer is assumed to be negligible so that $mA_1=nA_2$. Since $m<n$ due to the higher voltage set up by the individual top cells, $A_1$ must be larger than $A_2$. When calculating the current density of an area de-coupled device the total area $A$ is used as reference area. It is assumed that the two layers of cells are separated by a transparent material with a refractive index equal to that of the cells. Figure \ref{fig:areadecoupled} shows a sketch of the routes of photon transport within an area de-coupled device. A top cell will receive photons from more than one bottom cell, but since all bottom cells are equal and operate at the same voltage, the top cells will receive a flux of luminescent photons which is independent of the division of the cells in the bottom layer. Correspondingly, the luminescent flux of photons received by the bottom cells is independent of the division of cells in the top layer. The $JV$-characteristic of the top layer will therefore be

\begin{equation}
\label{eq:IVVMtop}
J_1=\frac{J_\mathrm{G1}A_\mathrm{1}}{A}-\frac{\left(1+n_\mathrm{r}^2\right)J_{0,1}A_\mathrm{1}}{A}\mathrm{e}^{\frac{qV}{mkT}}+\frac{n_\mathrm{r}^2J_{0,1}A_\mathrm{1}}{A}\mathrm{e}^{\frac{qV}{nkT}} =\frac{J_\mathrm{G1}}{m}-\frac{\left(1+n_\mathrm{r}^2\right)J_{0,1}}{m}\mathrm{e}^{\frac{qV}{mkT}}+\frac{n_\mathrm{r}^2J_{0,1}}{m}\mathrm{e}^{\frac{qV}{nkT}},
\end{equation}
where $V$ is the voltage set up by the entire device such that each individual cell in the top layer set up the voltage $V/m$ and each cell in the bottom layer set up a voltage $V/n$. The first term in Eq. (\ref{eq:IVVMtop}) is the generation current density associated with external photons, the second term is the recombination current density and the last term takes into account the generation current density associated with luminescence from the bottom layer. Correspondingly, the $JV$-characteristic of the bottom layer is given by

\begin{equation}
\label{eq:IVVMbot}
J_2=\frac{J_\mathrm{G2}}{n}-\frac{J_{0,2}+n_\mathrm{r}^2J_{0,1}}{n}\mathrm{e}^{\frac{qV}{nkT}}+ \frac{n_\mathrm{r}^2J_{0,1}}{n}\mathrm{e}^{\frac{qV}{mkT}}.
\end{equation}
The total current delivered by the area de-coupled tandem module is the sum of the currents from the two layers, i.e.

\begin{equation}
\label{eq:ADCJV}
J=J_1+J_2=\frac{J_\mathrm{G1}}{m}+\frac{J_\mathrm{G2}}{n}-J_{0,1}\left[\frac{1}{m}+n_\mathrm{r}^2\left(\frac{1}{m}-\frac{1}{n}\right)\right]\mathrm{e}^{\frac{qV}{mkT}}-\left[\frac{J_{0,2}}{n}+n_\mathrm{r}^2J_{0,1}\left( \frac{1}{n}-\frac{1}{m} \right)\right]\mathrm{e}^{\frac{qV}{nkT}}.
\end{equation}

As argued in section \ref{sec:multi}, the luminescence from the bottom cell to the top cell is usually negligible. Applying this approximation to Eq. (\ref{eq:IVVMtop}) allows the expression
\begin{equation}
\label{eq:VMPPvm1}
V_\mathrm{1,mpp}=\frac{mkT}{q}\left(\mathrm{W}\left[\frac{J_\mathrm{G1}}{\left(1+n_\mathrm{r}^2\right)J_\mathrm{0,1}}\mathrm{e} \right]-1 \right)
\end{equation}
to be established for the maximum power point of the top layer. This is $m$ times the voltage at the maximum power point for a stack of individually operated cells from Eq. (\ref{eq:Vmpp_top}). Similarly, the voltage at the maximum power point for the bottom layer is $n$ times the voltage given by Eqs. (\ref{eq:Vmpp_i}) and (\ref{eq:Zmpp_N}) with $N=2$. In an optimally designed device, the maximum power points of the two layers are equal. The optimal ratio of the number of top cells to the number of bottom cells is therefore
\begin{equation}
\frac{m}{n}=\frac{\mathrm{W}\left[\frac{J_\mathrm{G2}+\frac{n_\mathrm{r}^2}{1+n_\mathrm{r}^2}\frac{J_\mathrm{G1}}{\mathrm{W}\left(Z_1\right)}}{J_{0,2}}\mathrm{e}\right]-1}{\mathrm{W}\left[ \frac{J_\mathrm{G1}}{\left(1+n_\mathrm{r}^2\right) J_{0,1})}\mathrm{e}\right]-1},
\end{equation} 
where $Z_1$ is given by (\ref{eq:Zmpp_top}). When luminescent coupling is not occurring, the optimal ratio of $m$ to $n$ simplifies considerably to

\begin{equation}
\label{eq:optimalratio}
\frac{m}{n}=\frac{\mathrm{W}\left(\frac{J_\mathrm{G,2}}{J_{0,2}}\mathrm{e}\right)-1}{\mathrm{W}\left(\frac{J_\mathrm{G,1}}{J_{0,1}}\mathrm{e}\right)-1}.
\end{equation}
A similar approach may be followed to find the optimal number of cells in additional layers.

\section{Concluding remarks}
A number of $JV$-characteristics and expressions for key device parameters were derived for a variety of multi-junction concepts. The derivation is carried out assuming fully absorbing cells which operate at the radiative limit. The only simplifications or additional assumptions introduced to derive the $JV$-characteristic (\ref{eq:IVCMRC}) of a series-connected tandem stack is using the Boltzmann approximation. To derive the expressions for the cell parameters of the multi-terminal stacks, as well as the optimal number of cells in voltage-matched configurations, it was necessary to assume that the luminescence emitted from one cell to another cell with a larger band gap is negligible. This assumption is valid for the most interesting combinations of band gaps provided that the number of cells is not very large. 

The results have the potential to be used as a framework to introduce non-ideal processes like non-radiative recombination, for example by introducing the external luminescence extraction efficiency as in Ref. \cite{Pusch2019}. Other design features that may be incorporated to expanded models includes thinning of the top cell for better current matching as described by Kurtz et al \cite{Kurtz1990} or the addition of layers with a low refractive index between cells to curtail the luminescent coupling, as suggested by Sheng et al. \cite{Sheng2015}.

\appendix

\begin{figure}[htbp]
\centering
\includegraphics[width=9cm]{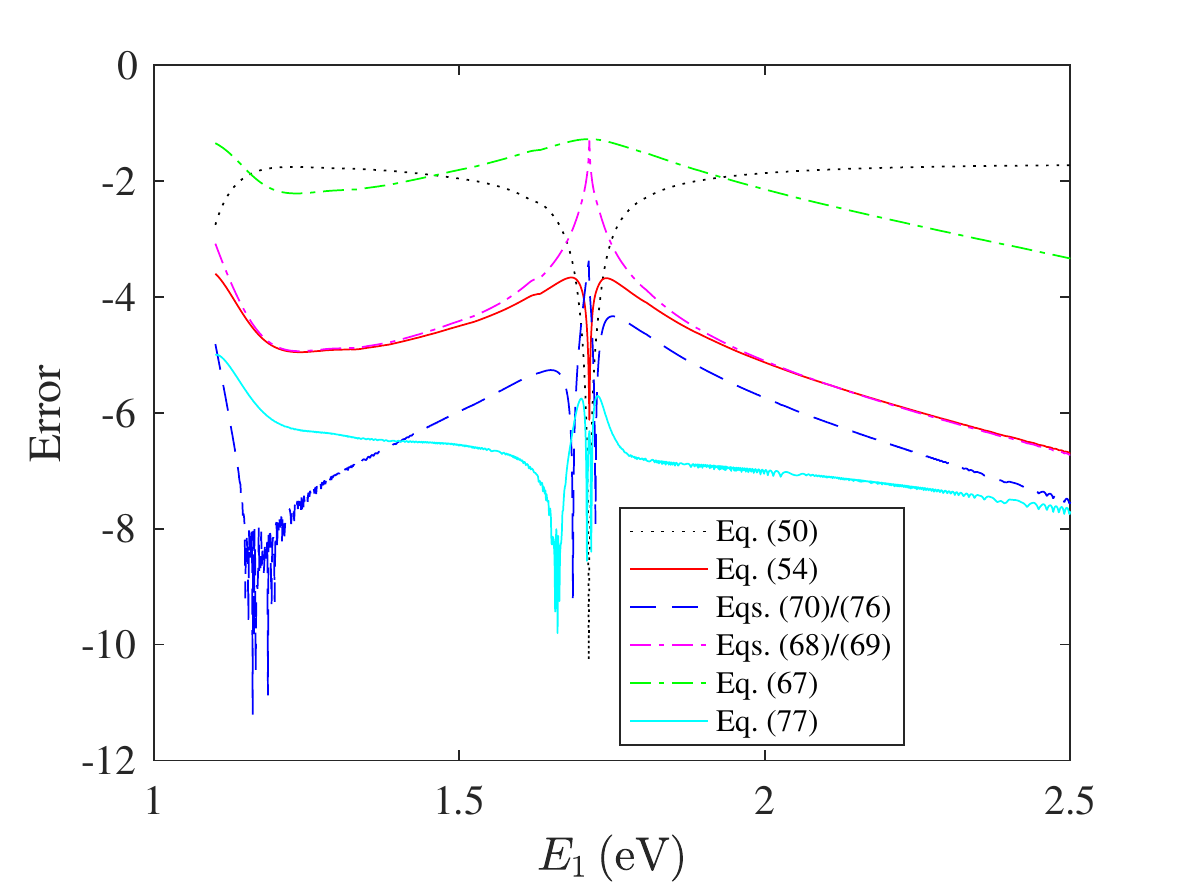}
\caption{The logarithm of the relative error of the maximum power density for stacks with a refractive index of 2.5 calculated using a variety of expressions. The band gap of the bottom cell is set to 1.1 eV and the AM1.5G spectrum has been used to calculate the generation current densities. $\Delta J_\mathrm{G}=0$ for $E_1=1.71\,\mathrm{eV}$.  \label{fig:linjeplott}}
\end{figure}

\begin{figure}[htbp]
\centering
\includegraphics[width=9cm]{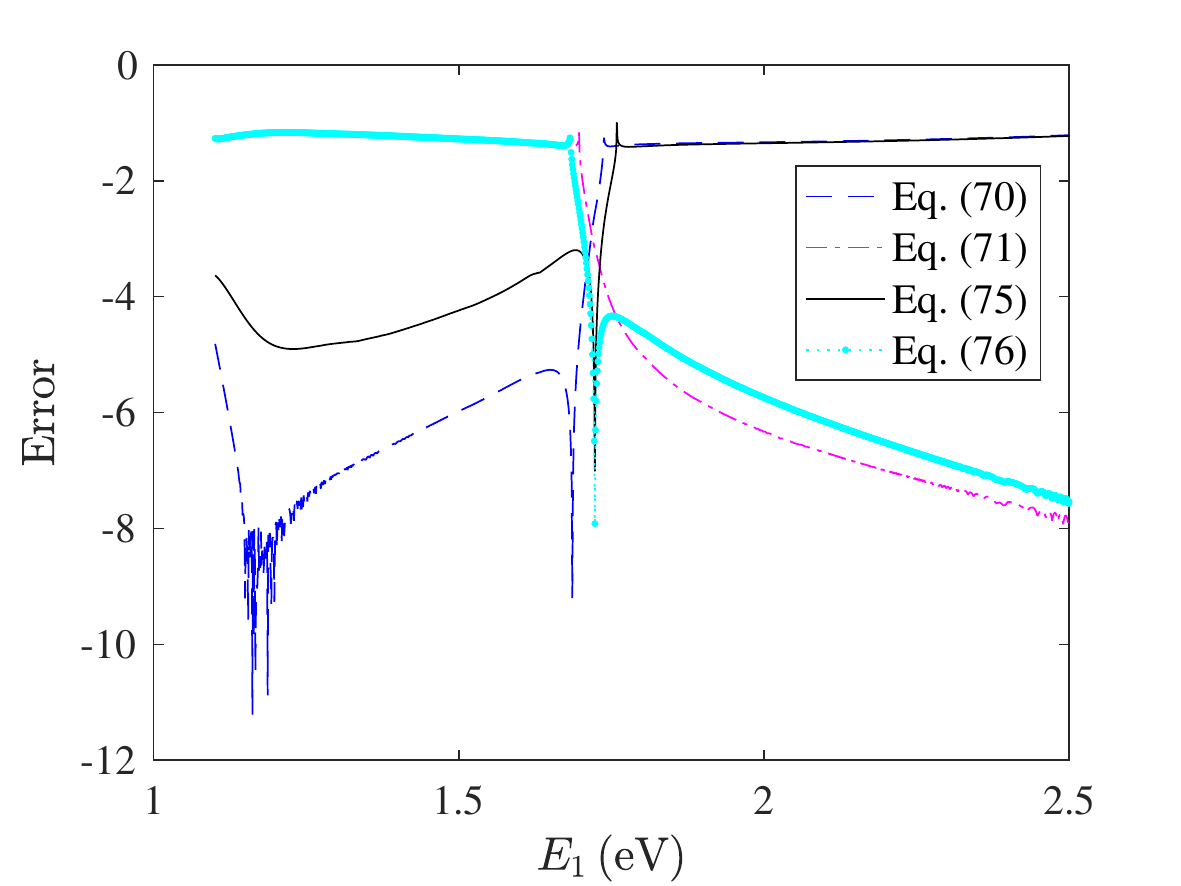}
\caption{The logarithm of the relative error of the maximum power density for stacks with a refractive index of 2.5 calculated using Eqs. (\ref{eq:Vmpp031}), (\ref{eq:Vmpp032}), (\ref{eq:Vmpp041}) and (\ref{eq:Vmpp042}). The band gap of the bottom cell is set to 1.1 eV and the AM1.5G spectrum has been used to calculate the generation current densities. $\Delta J_\mathrm{G}=0$ for $E_1=1.71\,\mathrm{eV}$.  \label{fig:linjeplott2}}
\end{figure}

This appendix explores the maximum power point of a series-connected tandem cell. Several expressions are derived or constructed, and their accuracy is investigated. The latter is quantified by the logarithm of the relative error, as explained in Sec. \ref{sec:mpp}, and plotted in Figs. \ref{fig:linjeplott} and \ref{fig:linjeplott2}. The examples shown in these figures are calculated with a refractive index of 2.5 and a bottom cell band gap of 1.1 eV. The AM1.5G spectrum is used.

The first step in exploring the maximum power point of a tandem stack is to rewrite Eq. (\ref{eq:Vmppcm}). Using the identities
\begin{equation}
\label{eq:ident1}
2\mathrm{W}\left(x\right)=\mathrm{W}\left(\frac{2x^2}{\mathrm{W}\left(x\right)} \right),
\end{equation}
\begin{equation}
\label{eq:ident2}
\mathrm{W}\left(x \right)+\mathrm{W}\left(y\right)=\mathrm{W}\left( xy\left(\frac{1}{\mathrm{W}\left(x \right)}+\frac{1}{\mathrm{W}\left(y \right)} \right)\right)
\end{equation}
and
\begin{equation}
\label{eq:ident1}
\mathrm{W}\left(-\mathrm{e}^{-1}\right)=-1,
\end{equation}
allows Eq. (\ref{eq:Vmppcm}) to be expressed as
\begin{equation}
\label{eq:Vmpprewrite}
\frac{qV_\mathrm{mpp}}{kT}+1=\mathrm{W}\left[\mathrm{e}^{-1}\tilde{Z}_1\tilde{Z}_2 \left( \frac{2}{\mathrm{W}\left(\sqrt{\tilde{Z}_1\tilde{Z}_2}\right)}-\frac{1}{\mathrm{W}\left(\sqrt{\tilde{Z}_1\tilde{Z}_2}\right)^2} \right)\right].
\end{equation}
This result will be used later.

Pusch et al. \cite{Pusch2019} utilized the fact that the cell receiving the largest photon flux will have a steep $JV$-characteristic around the maximum power point of the stack, as long as $\Delta J_\mathrm{G}$ is not too small. For $\Delta J_\mathrm{G}>0$, $V_1$ can then be treated as a constant while finding the optimal value of $V_2$ and vice versa. They arrived at an expression\footnote{Eq. (10) in Ref. \cite{Pusch2019}.} equivalent to

\begin{equation}
\label{eq:Vmpp001}
\frac{qV_\mathrm{mpp}}{kT}+1=\mathrm{W}\left[\tilde{Z}_2\mathrm{e}^\frac{qV_1}{kT}\right],
\end{equation}
if $\Delta J_\mathrm{G}>0$, and
\begin{equation}
\label{eq:Vmpp002}
\frac{qV_\mathrm{mpp}}{kT}+1=\mathrm{W}\left[\tilde{Z}_1\mathrm{e}^\frac{qV_2}{kT}\right],
\end{equation}
if $\Delta J_\mathrm{G}<0$, for the maximum power point of the stack. If $\vert\Delta J_\mathrm{G}\vert$ is large they suggested inserting the respective open circuit voltages for $V_1$ and $V_2$, which gives
 
\begin{equation}
\label{Vmpp010}
\frac{qV_\mathrm{mpp}}{kT}+1=\mathrm{W}\left[\mathrm{e}^{-1}\tilde{Z}_1\tilde{Z}_2\right],
\end{equation}
regardless of the polarity of $\Delta J_\mathrm{G}$. Although this gives a fair estimate of the maximum power for many combinations of band gaps, the relative error is rather large, as shown in Fig. \ref{fig:linjeplott}, particularly when $\vert\Delta J_\mathrm{G}\vert$ is small. One possible improvement is to insert the short circuit voltage of the cell receiving the largest photon flux into (\ref{eq:Vmpp001}) or (\ref{eq:Vmpp001}). This gives

\begin{equation}
\label{eq:Vmpp021}
\frac{qV_\mathrm{mpp}}{kT}+1=\mathrm{W}\left[\mathrm{e}^{-1}\tilde{Z}_1\tilde{Z}_2\frac{\left(1-T^+\right)\Delta J_\mathrm{G}}{J_\mathrm{G1}-T_{2\rightarrow 1}\Delta J_\mathrm{G}}\right],
\end{equation}
when $\Delta J_\mathrm{G}$ is positive, and
\begin{equation}
\label{eq:Vmpp022}
\frac{qV_\mathrm{mpp}}{kT}+1=\mathrm{W}\left[\mathrm{e}^{-1}\tilde{Z}_1\tilde{Z}_2\frac{\left(T^+-1\right)\Delta J_\mathrm{G}}{J_\mathrm{G2}+T_{1\rightarrow 2}\Delta J_\mathrm{G}}\right],
\end{equation}
when $\Delta J_\mathrm{G}$ is negative. Using the short circuit voltage instead of the open circuit voltage reduces the error significantly for most combinations of band gaps, but it is still rather large when $\vert\Delta J_\mathrm{G}\vert$ is small, as shown in Fig. \ref{fig:linjeplott}. Another option which was suggested in Ref. \cite{Pusch2019} is to find the individual maximum power point of the cell receiving the smallest photon flux and using the corresponding current to determine the voltage of the other cell. This gives
 
\begin{equation}
\label{eq:Vmpp031}
\frac{qV_\mathrm{mpp}}{kT}+1=\mathrm{W}\left[\mathrm{e}^{-1}\tilde{Z}_1\tilde{Z}_2\frac{J_\mathrm{G2}+T_{1\rightarrow 2}\Delta J_\mathrm{G}}{J_\mathrm{G1}-T_{2\rightarrow 1}\Delta J_\mathrm{G}}\left(\frac{\left(1-T^+\right)\Delta J_\mathrm{G}}{J_\mathrm{G2}+T_{1\rightarrow 2}\Delta J_\mathrm{G}}+\frac{1}{\mathrm{W}\left(\tilde{Z}_2\right)}\right)\right],
\end{equation}
if $\Delta J_\mathrm{G}$ is positive, and
\begin{equation}
\label{eq:Vmpp032}
\frac{qV_\mathrm{mpp}}{kT}+1=\mathrm{W}\left[\mathrm{e}^{-1}\tilde{Z}_1\tilde{Z}_2\frac{J_\mathrm{G1}-T_{2\rightarrow 1}\Delta J_\mathrm{G}}{J_\mathrm{G2}+T_{1\rightarrow 2}\Delta J_\mathrm{G}}\left(\frac{\left(T^+-1\right)\Delta J_\mathrm{G}}{J_\mathrm{G1}-T_{2\rightarrow 1}\Delta J_\mathrm{G}}+\frac{1}{\mathrm{W}\left(\tilde{Z}_1\right)}\right)\right],
\end{equation}
if $\Delta J_\mathrm{G}$ is negative. The logarithm of the relative error of these approximations are shown in Fig. \ref{fig:linjeplott2}. 

In addition to Eq. (\ref{eq:Vmppcm}), which is exact for $\Delta J_\mathrm{G}=0$, it is possible to find an exact solution for the maximum power point for another special case. That is when the two cells deliver the same current density at their individual maximum power points. The $V_\mathrm{mpp}$ is then the sum of the optimal voltages of the two individual cells, which gives 
\begin{equation}
\label{eq:equalJmpp}
V_\mathrm{mpp}=\frac{kT}{q}\left(\mathrm{W}\left(\tilde{Z}_1\right)+ \mathrm{W}\left(\tilde{Z}_2\right)-2\right).
\end{equation}
Using the identities (\ref{eq:ident2}) and (\ref{eq:ident1}) allows Eq. (\ref{eq:equalJmpp}) to be rewritten as
\begin{equation}
\label{eq:equalJmpp2}
\frac{qV_\mathrm{mpp}}{kT}+1=\mathrm{W}\left[\mathrm{e}^{-1}\tilde{Z}_1\tilde{Z}_2\left(\frac{1}{\mathrm{W}\left(\tilde{Z}_1\right)}+ \frac{1}{\mathrm{W}\left(\tilde{Z}_2\right)}-\frac{1}{\mathrm{W}\left(\tilde{Z}_1\right)\mathrm{W}\left(\tilde{Z}_2\right)}\right)\right].
\end{equation}
For this special case we also have
\begin{equation}
\label{eq:equalcurrent}
\Delta J_\mathrm{G}\left(1-T^+\right)=\frac{J_\mathrm{G,1}-T_{1\rightarrow 2}\Delta J_\mathrm{G}}{\mathrm{W}(\tilde{Z}_1)}-\frac{J_\mathrm{G2}+T_{2\rightarrow 1}\Delta J_\mathrm{G}}{\mathrm{W}(\tilde{Z}_2)},
\end{equation}
which is found by equating $J_1$ to $J_2$ at $V_\mathrm{mpp}$. This can be used to substitute for either $\tilde{Z}_1$ or $\tilde{Z}_2$ in (\ref{eq:equalJmpp2}), which gives
\begin{equation}
\label{eq:Vmpp041}
\frac{qV_\mathrm{mpp}}{kT}+1=\mathrm{W}\left[\mathrm{e}^{-1}\tilde{Z}_1\tilde{Z}_2\frac{J_\mathrm{G2}+T_{1\rightarrow 2}\Delta J_\mathrm{G}}{J_\mathrm{G1}-T_{2\rightarrow 1}\Delta J_\mathrm{G}}\left( \frac{\left(1-T^+\right)\Delta J_\mathrm{G}}{J_\mathrm{G2}+T_{1\rightarrow 2}\Delta J_\mathrm{G}} +\frac{2}{\mathrm{W}(\tilde{Z}_2)}-\frac{1}{\mathrm{W}(\tilde{Z}_2)^2}\right)\right]
\end{equation}
and
\begin{equation}
\label{eq:Vmpp042}
\frac{qV_\mathrm{mpp}}{kT}+1=\mathrm{W}\left[\mathrm{e}^{-1}\tilde{Z}_1\tilde{Z}_2\frac{J_\mathrm{G1}-T_{2\rightarrow 1}\Delta J_\mathrm{G}}{J_\mathrm{G2}+T_{1\rightarrow 2}\Delta J_\mathrm{G}}\left( \frac{\left(T^+-1\right)\Delta J_\mathrm{G}}{J_\mathrm{G1}-T_{2\rightarrow 1}\Delta J_\mathrm{G}} +\frac{2}{\mathrm{W}(\tilde{Z}_1)}-\frac{1}{\mathrm{W}(\tilde{Z}_1)^2}\right)\right],
\end{equation}
respectively.
The logarithm of the relative error of the maximum power density calculated using Eqs. (\ref{eq:Vmpp041}) and (\ref{eq:Vmpp042}) are also plotted in Fig. \ref{fig:linjeplott2}. It is seen that a combination of (\ref{eq:Vmpp031}), for positive $\Delta J_\mathrm{G}$, and (\ref{eq:Vmpp042}), for negative $\Delta J_\mathrm{G}$, gives the smallest maximum error across the various top cell band gaps. This combination has therefore been plotted in Fig. \ref{fig:linjeplott} to be compared with the error of other expressions. In Ref. \cite{Pusch2019}, further expressions were suggested to get good accuracy for small values of $\vert\Delta J_\mathrm{G}\vert$, but these are not explored further here.

The similarities between Eqs. (\ref{eq:Vmpprewrite}), (\ref{eq:Vmpp031}) and (\ref{eq:Vmpp042}) motivates the construction of Eq. (\ref{eq:VmppCMTS}). One important advantage of Eq. (\ref{eq:VmppCMTS}) is that this single expression gives a small error for all the band gap combinations examined in Figs. \ref{fig:error}, regardless of the value of $\Delta J_\mathrm{G}$. As is seen from Fig. \ref{fig:linjeplott}, Eq. (\ref{eq:VmppCMTS}) also has a smaller maximum relative error than any of the other expressions presented above. One weakness of (\ref{eq:Vmpprewrite}) is that it is not the most accurate expression for larger values of $\vert\Delta J_\mathrm{G}\vert$. 
\begin{figure}[htbp]
\centering
\includegraphics[width=9cm]{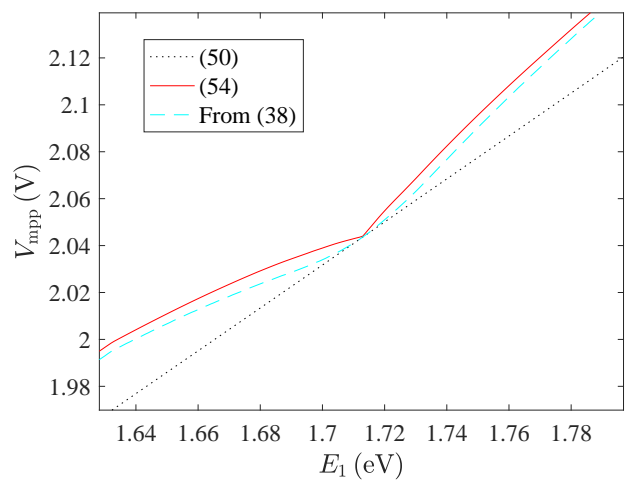}
\caption{The $V_\mathrm{mpp}$ calculated for the AM1.5G spectrum with a bottom cell band gap of 1.1 eV as a function of the top cell band gap. $\Delta J_\mathrm{G}$ is zero for a top cell band gap of 1.71 eV. The dashed line is the $V_\mathrm{mpp}$ found numerically from the $JV$-characteristic (\ref{eq:IVCMRC}).  \label{fig:Vmpp}}
\end{figure}

\begin{figure}[htbp]
\centering
\includegraphics[width=9cm]{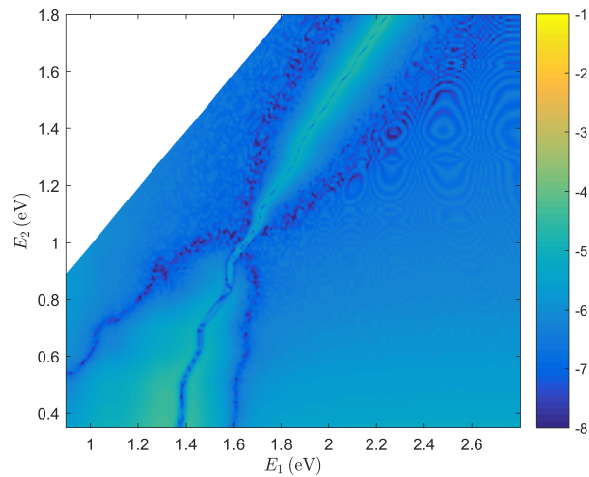}
\caption{The logarithm of the relative error of the maximum power density for stacks without luminescent coupling calculated using Eq. (\ref{eq:VmppCMTSimpr}.) The AM1.5G spectrum has been used. The largest value of -4.15 is found for a device with band gaps of 0.35 and 1.29 eV. \label{fig:imprmap}}
\end{figure}

It is possible to improve (\ref{eq:VmppCMTS}) by subtracting a term containing $\mathrm{W}\big(\sqrt{\tilde{Z}_1\tilde{Z}_2}\big)^{-2}$, but this only gives a slight reduction in the maximum error. Another way of improving (\ref{eq:VmppCMTS}) is motivated by the plot of the $V_\mathrm{mpp}$ in Fig. \ref{fig:Vmpp}. Here it can be seen that the value of the $V_\mathrm{mpp}$ calculated by (\ref{eq:VmppCMTS}) changes too fast as $\vert \Delta J_\mathrm{G}\vert$ grows from zero. This motivates a damping of the term containing $\Delta J_\mathrm{G}$. At the expense of increased complexity, the expression 
\begin{equation}
\label{eq:VmppCMTSimpr}
\frac{qV_\mathrm{mpp}}{kT}+1\approx\mathrm{W}\left[\frac{J_{\mathrm{sc}}^2\mathrm{e}}{(1-T^+)^2\tilde{J}_0^2}\left((1-T^+)\frac{\vert\Delta J_\mathrm{G}\vert}{J_\mathrm{sc}}\mathrm{e}^\frac{-J_\mathrm{sc}}{40(1-T^+)\vert \Delta J_\mathrm{G}\vert}+\frac{2}{\mathrm{W}\left(\sqrt{\tilde{Z}_1\tilde{Z}_2}\right)}\right)\right]
\end{equation}
gives the $V_\mathrm{mpp}$ with only a very small error. The factor of 40 is fitted to minimize the error for the band gap combinations in Fig. \ref{fig:linjeplott}. The logarithm of the relative error using this last expression is included in Figs. \ref{fig:linjeplott} and explored further in \ref{fig:imprmap}. The latter figure shows a map equivalent to that in Fig. \ref{fig:error}. The largest error in the map is -4.15, which corresponds to $7\cdot10^{-3}\,\%$. 

\bibliographystyle{IEEEtran}
\bibliography{bibliography}

\end{document}